# Electron Paramagnetic Resonance Studies of KYb(MoO$_4$)$_2$

K.G. Dergachev, M.I. Kobets, E.N. Khatsko

Institute for Low Temperature Physics National Academy of Science of Ukraine

47 Lenin Ave.61103 Kharkov, Ukraine

khatsko@ilt.kharkov.ua

The electron paramagnetic resonance investigations of the magnetically concentrated crystal of KYb(MoO$_4$)$_2$ have been performed. The main value of g-factors along principal local axes was determinate. The two nonequivalent Yb$^{3+}$ centers were found in *ac*-plane. It is shown that local symmetry of Yb$^{3+}$ ion in *ac*-plane is not higher than rhombic. Some peculiarities in the frequency-field dependence of an absorption line was found in **H**|| **b** orientation.

A permanent interest to studying of a double alkali-rare-earth molybdates and tungstates is associated with the fact that many of them are laser materials, ferroelectrics, piezoelectrics, and ferroelastics. In addition, many peculiarities of electronic and magnetic subsystems, such as strong electron-phonon interaction, a layered structure, a nonharmonism of interparticles interaction are inherent characteristic of the HTS compounds as well. Structural phase transitions in these compounds occur at low temperatures because of a cooperative Jahn-Teller effect. As a consequence, studying of this family of magnetically concentrated crystals with low symmetry of structure, as well as searching for field induced structural phase transition in these compounds is appropriate and topical problem wich offers possibilities of detection new properties and phenomena typical only for low symmetric crystals.

A representative of this family the single crystal of KYb(MoO$_4$)$_2$ was investigate by EPR method in 50-120 GHz frequency range at helium temperature. The layered potassium–ytterbium double molybdat have a rhombic symmetry of the crystal structure (space group $D_{2h}^{14}$ – *Pbcn*) and contain four formulae units per elementary cell. The cell parameters are *a* = 5,06 Å; *b* = 7,85 Å; *c* = 18,32 Å [1]. The Yb$^{3+}$ ions are surrounded by 8 oxygen's forming a distorted antiprism polyhedra (as the EPR indicate). The shortest distance between Yb$^{3+}$ - Yb$^3$ in chain situated along *c*-axis is 3.975 Å. The crystal structure fragment is shown in Fig.1. All crystals of the double molybdat family have a layered structure and split easily on thin plates. Typically **b**-axis is perpendicular to splitting plane. The crystal under investigation was obtained by spontaneous crystallizing method from solution in melt. A crystallization process takes place in massive platinum crucible at 1000º C. The temperature was stabilized. The qualitative single crystals were crystallized on the surface of a solution and after cooling can be easily separate from the solvent.

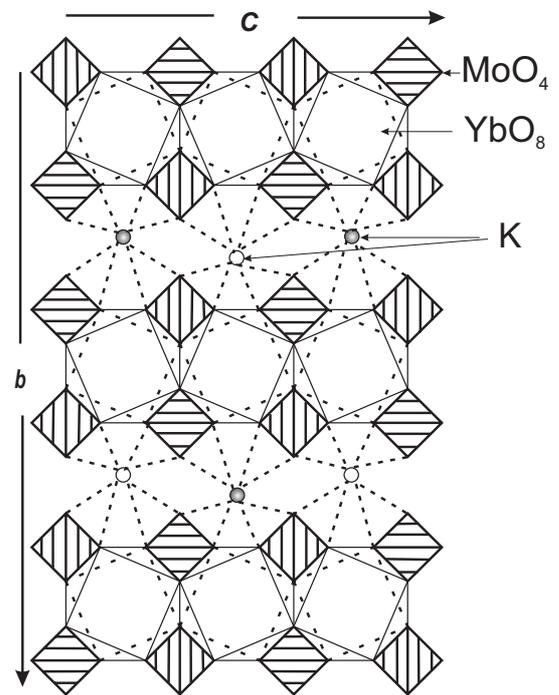

FIG. View of a fragment of KYb(MoO$_4$)$_2$ single crystal structure

The ground state of f-electrons of Yb$^{3+}$ ion in KYb(MoO$_4$)$_2$ is - $^2F_{7/2}$. It is spitted up to single Kramers doublets. So the lowest state can be described by spin-Hamiltonian for effective spin S=½ with anisotropic g-factor without a fine structure of spectrum. From optical data $Yb^{3+}$ ion have a simple structure of energy levels. The energy gap ΔE between ground state and first exited state of $^2F_{7/2}$ multiplet is big enough and at temperature 4.2 K averages amount in order of $\Delta E = 150 – 200$ cm$^{-1}$. Thus at low temperature EPR spectrum is formed mainly by the lowest Kramers doublet. Along with this the basic term $^2F_{7/2}$ of $Yb^3$ ion have nonzero orbital and spin momentum, thus all peculiarities of EPR spectrum connected with anisotropic spin-spin interaction are retained, but mechanisms caused by mixing of exited levels will be very weaken.

Previously [2] EPR spectrum in $KYb(MoO_4)_2$ have been investigated at low frequency (f=10 GHz). Because at this frequency the resolution is low enough one can see only a distorted line they could only estimate g-factors in *ac*-plane.

For this reason we study EPR spectra of $KYb(MoO_4)_2$ at high frequency in 50-120 GHz range at temperature 4.2 K in the strong field condition. Angular and frequency-field dependences of EPR spectrum was studied in perpendicular polarization H⊥h. Two geometrically nonequivalent $Yb^{3+}$ ion centres in $KYb(MoO_4)_2$ lattice was found from studies of angular dependences EPR line position in ac-plane. Fig 2 demonstrates a typical absorption spectrum. The line width of EPR line is $\Delta H \approx 650$ Oe at magnetic field orientation along local c'-axis. Fig. 3 presents the angular dependences of EPR line position in *ac*-plane. The local axes *a'* and *c'* of centers a rotated in relation to crystalline axes *a* and *c* to the angle $\varphi = \pm (34\pm0,2^o)$. An existence of two geometrically nonequivalent centers is in accordance with crystalline structure of the double molybdate $KYb(MoO_4)_2$. Measured main values of g-tensor for both centers are the same, suggesting that centers are equivalent. The values of g-tensor components are $g_{a'} =1,8 \pm 0,05; g_{c'} = 6,34 \pm 0,05$.

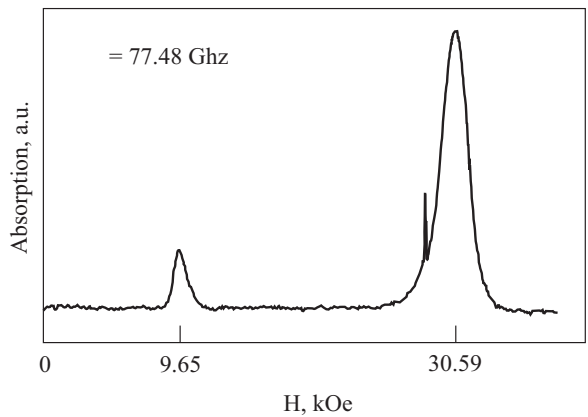

FIG.2 Typical EPR spectrum of $KYb(MoO_4)_2$

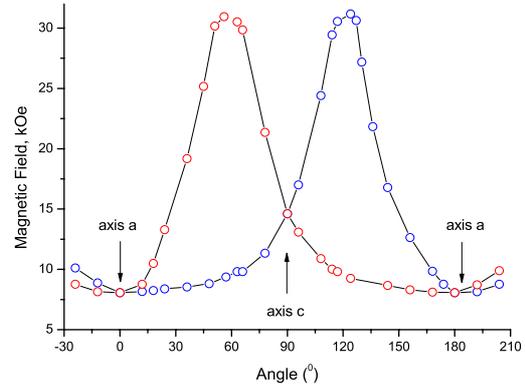

FIG.3. The angular dependence of the EPR spectrum of $KYb(MoO_4)_2$ in *ac*-plane (T=4.2 K)

In addition, the strong dependence of the line intensity on magnetic field direction was observed. The intensity is at a minimum when magnetic field is directed along H||*c'* and it is at a maximum when H||*a'*. This intensity anisotropy can be explained as follows. We use in experiment perpendicular orientation of persistent magnetic field H and ultra high frequency(UHF) magnetic field h (H⊥h). In this case at H||*c'* orientation the UHF field h directed along axis with minimal magnetic susceptibility, so the EPR line absorption is minimal. In case H||*a'* h coincide with maximum susceptibility direction, so absorption is maximal.

Only one symmetric absorption line with g-factor $g = 1.48 \pm 0.05$ was observed for H||*b* at frequency 56.85 GHz as axes of both centers coincide in this case. The line width was $\Delta H_b \approx 3.5$ kOe. The frequency-field dependence of EPR spectrum was investigated at 4.2 K for more precisely defining of g-factor value. This dependence shows that the lowest ground state is Kramers doublet. But as the frequency increases an essentially change in EPR spectrum take

place. A smooth splitting of resonant line into two components with equal intensity was observed at severe orientation H||b. The following frequency increasing leads to stronger splitting of resonant line. This behavior can not be explained by monoclinic component in Hamiltonian and up to now we have not rigorous theoretical explanation for observed spectrum transformation. It is known from literature that the next exited level in this compound is situated at distance $\Delta E \approx 150$ cm$^{-1}$ so dispushing of the ground state and the first exited one can not be occur. It remains to suppose that probably we see a dynamic interaction between electron branch and low frequency phonon branch. But such an assumption needs additionally studying.

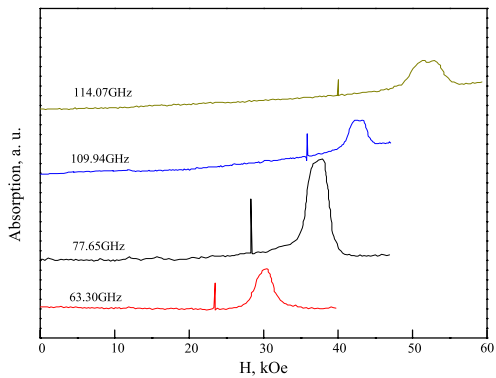

FIG.4. The frequency dependence of absorption line shape in H||b direction.

The line form is a Lorenz type. An anisotropic part of spin-spin interactions is mainly determined by magnetic dipole-dipole interaction, which forms resonant line. The energy of this interaction is considerably less than in other alkali-rare earth molybdates, as $KDy(MoO_4)_2$, $KEr(MoO_4)_2$ in connection with the lesser value of $Yb^{3+}$ ion magnetic moment. The value of dipole-dipole interaction under our estimation is $E_{dd}\sim 0,1$ cm$^{-1}$. It is possible that direct relaxation processes are of considerable importance. They are characterized by spin-phonon binding enhancement when external magnetic field rises. This may result in moderate splitting of a resonant line as the frequency increases.

In the case when the Zeemann splitting energy is not too high in relation to the distance to the first exited doublet (this is the $Yb^{3+}$ in $KYb(MoO_4)_2$ case) and the nonlinear in H effects can be neglected the next spin Hamiltonian can be used for each center:

$$H_{c'} = g\beta HS$$

Where S is effective spin S=1/2; g – Lande factor; β- Bohr magneton.

The sample temperature under spectra measurements was over spin-spin interaction energy, so no peculiarities, connected with spin polarization (i.e. magnetization of the crystal) was not observed.

It is known [3,4] that in double alkali-rare-earth molybdates structural 1-st kind type phase transitions induced by magnetic field was observed. The attempts of theoretical description these transitions [5,6] are based on a model of the strong spin-phonon coupling end existence of quasi-degenerated ground state. In $KYb(MoO_4)_2$, where exited state is separated by 150-200 cm$^{-1}$ from ground one and spin-phonon coupling is depressed, we do not observe any anomalies pointed to existence of such transitions. This is an indirect argument in favour of chosen theoretical models.

In summary we can formulate the experimental results.

1. For magnetically concentrated crystal of $KYb(MoO_4)_2$ the main value of g-factors along principal local axes was determinate.

2. The two nonequivalent $Yb^{3+}$ centers were found in *ac*-plane.

3. It is shown that local symmetry of $Yb^{3+}$ ion in *ac*-plane is not higher than rhombic.

4. Some peculiarities in the frequency-field dependence of an absorption line was found in **H|| *b*** orientation.